\documentclass[letterpaper,twocolumn,english,aps,prx,floatfix,superscriptaddress]{revtex4}
\usepackage[T1]{fontenc}
\usepackage[latin9]{inputenc}
\usepackage{color}
\usepackage{amsmath}
\usepackage{amssymb}
\usepackage{graphicx}
\usepackage{esint}

\makeatletter

\pdfpageheight\paperheight
\pdfpagewidth\paperwidth

\@ifundefined{textcolor}{}
{%
 \definecolor{BLACK}{gray}{0}
 \definecolor{WHITE}{gray}{1}
 \definecolor{RED}{rgb}{1,0,0}
 \definecolor{GREEN}{rgb}{0,1,0}
 \definecolor{BLUE}{rgb}{0,0,1}
 \definecolor{CYAN}{cmyk}{1,0,0,0}
 \definecolor{MAGENTA}{cmyk}{0,1,0,0}
 \definecolor{YELLOW}{cmyk}{0,0,1,0}
}

\renewcommand{\[}{\begin{equation}}
\renewcommand{\]}{\end{equation}}

\usepackage{babel}

\usepackage{babel}

\usepackage{babel}

\makeatother

\usepackage{babel}
\begin{document}
\global\long\def\avg#1{\langle#1\rangle}

\global\long\def\p{\prime}

\global\long\def\dg{\dagger}

\global\long\def\ket#1{|#1\rangle}

\global\long\def\bra#1{\langle#1|}

\global\long\def\proj#1#2{|#1\rangle\langle#2|}

\global\long\def\inner#1#2{\langle#1|#2\rangle}

\global\long\def\tr{\mathrm{tr}}

\global\long\def\pd#1#2{\frac{\partial#1}{\partial#2}}

\global\long\def\spd#1#2{\frac{\partial^{2}#1}{\partial#2^{2}}}

\global\long\def\der#1#2{\frac{d#1}{d#2}}

\global\long\def\im{\imath}

\global\long\def\As{{^{\sharp}}\hspace{-1mm}\mathcal{A}}

\global\long\def\Fs{{^{\sharp}}\hspace{-0.7mm}\mathcal{F}}

\global\long\def\Es{{^{\sharp}}\hspace{-0.5mm}\mathcal{E}}

\global\long\def\Fd{{^{\sharp}}\hspace{-0.7mm}\mathcal{F}_{\delta}}

\global\long\def\S{\mathcal{S}}

\global\long\def\A{\mathcal{A}}

\global\long\def\F{\mathcal{F}}

\global\long\def\E{\mathcal{E}}

\global\long\def\O{\mathcal{O}}

\global\long\def\SgF{\S d\F}

\global\long\def\SgEF{\S d\left(\E/\F\right)}

\global\long\def\U{\mathcal{U}}

\global\long\def\V{\mathcal{V}}

\global\long\def\H{\mathbf{H}}

\global\long\def\SO{\Pi_{\S}}

\global\long\def\PO{\hat{\Pi}_{\S}}

\global\long\def\SSH{\tilde{\Pi}_{\S}}

\global\long\def\EO{\Upsilon_{k}}

\global\long\def\ESH{\Omega_{k}}

\global\long\def\HSF{\mathbf{H}_{\S\F}}

\global\long\def\HSEF{\mathbf{H}_{\S\E/\F}}

\global\long\def\HS{\mathbf{H}_{\S}}

\global\long\def\ES{H_{\S}(t)}

\global\long\def\ESo{H_{\S}(0)}

\global\long\def\EgF{H_{\SgF} (t)}

\global\long\def\EgE{H_{\S d\E}(t)}

\global\long\def\EgEF{H_{\SgEF} (t)}

\global\long\def\EF{H_{\F}(t)}

\global\long\def\EFo{H_{\F}(0)}

\global\long\def\ESF{H_{\S\F}(t)}

\global\long\def\ESEF{H_{\S\E/\F}(t)}

\global\long\def\ESSEF{H_{\tilde{\S}\S\E/\F}(t)}

\global\long\def\EEFo{H_{\E/\F}(0)}

\global\long\def\EEF{H_{\E/\F}(t)}

\global\long\def\MI{I\left(\S:\F\right)}

\global\long\def\aMI{\left\langle \MI\right\rangle _{\Fs}}

\global\long\def\BS{\Pi_{\S} }

\global\long\def\PB{\hat{\Pi}_{\S} }

\global\long\def\QD{\mathcal{D}\left(\Pi_{\S}:\F\right)}

\global\long\def\QDp{\mathcal{D}\left(\PB:\F\right)}

\global\long\def\JI{J\left(\Pi_{\S}:\F\right)}

\global\long\def\CI{H\left(\F\left|\Pi_{\S}\right.\right)}

\global\long\def\CIp{H\left(\F\left|\PB\right.\right)}

\global\long\def\CS{\rho_{\F\left|s\right.}}

\global\long\def\CSu{\tilde{\rho}_{\F\left|s\right.}}

\global\long\def\CSp{\rho_{\F\left|\hat{s}\right.}}

\global\long\def\CEF{H_{\F\left|s\right.}}

\global\long\def\CEFp{H_{\F\left|\hat{s}\right.}}

\global\long\def\psiz{\ket{\psi_{\E\left|0\right.\hspace{-0.4mm}}}}

\global\long\def\psio{\ket{\psi_{\E\left|1\right.\hspace{-0.4mm}}}}

\global\long\def\psiinner{\inner{\psi_{\E\left|0\right.\hspace{-0.4mm}}}{\psi_{\E\left|1\right.\hspace{-0.4mm}}}}

\global\long\def\QDz{\boldsymbol{\delta}\left(\S:\F\right)_{\left\{  \sigma_{\S}^{z}\right\}  }}

\global\long\def\NQD{\bar{\boldsymbol{\delta}}\left(\S:\F\right)_{\BS}}

\global\long\def\EFS{H_{\F\left| \BS\right. }(t)}

\global\long\def\EFSM{H_{\F\left| \left\{  \ket m\right\}  \right. }(t)}

\global\long\def\Hol{\chi\left(\Pi_{\S}:\F\right)}

\global\long\def\Holp{\chi\left(\PB:\F\right)}

\global\long\def\ch{\raisebox{0.5ex}{\mbox{\ensuremath{\chi}}}_{\mathrm{Pointer}}}

\global\long\def\rhoS{\rho_{\S}(t)}

\global\long\def\rhoSo{\rho_{\S}(0)}

\global\long\def\rhoSF{\rho_{\S\F} (t)}

\global\long\def\rhoSgEF{\rho_{\SgEF} (t)}

\global\long\def\rhoSgF{\rho_{\SgF} (t)}

\global\long\def\rhoF{\rho_{\F}(t)}

\global\long\def\rhoFp{\rho_{\F}(\pi/2)}

\global\long\def\LE{\Lambda_{\E}(t)}

\global\long\def\LEc{\Lambda_{\E}^{\star}(t)}

\global\long\def\LEij{\Lambda_{\E}^{ij}(t)}

\global\long\def\LF{\Lambda_{\F}(t)}

\global\long\def\LFij{\Lambda_{\F}^{ij} (t)}

\global\long\def\LFc{\Lambda_{\F}^{\star}(t)}

\global\long\def\LEF{\Lambda_{\E/\F} (t)}

\global\long\def\LEFij{\Lambda_{\E/\F}^{ij}(t)}

\global\long\def\LEFc{\Lambda_{\E/\F}^{\star}(t)}

\global\long\def\Lkij{\Lambda_{k}^{ij}(t)}

\global\long\def\Hb{H}

\global\long\def\kE{\kappa_{\E}(t)}

\global\long\def\kEF{\kappa_{\E/\F}(t)}

\global\long\def\kF{\kappa_{\F}(t)}

\global\long\def\ts{t=\pi/2}

\global\long\def\QCB{\bar{\xi}_{QCB}}

\global\long\def\mc#1{\mathcal{#1}}

\global\long\def\MD{\lambda}

\global\long\def\up{\uparrow}

\global\long\def\down{\downarrow}

\global\long\def\Cku{\rho_{k\left|\up\right.}}

\global\long\def\Ckd{\rho_{k\left|\down\right.}}

\global\long\def\f{\mathcal{J}}

\global\long\def\onlinecite#1{\cite{#1}}

\title{Amplification, Decoherence, and the Acquisition of Information by
Spin Environments}

\author{Michael Zwolak}

\email{mpzwolak@gmail.com}

\address{Department of Physics, Oregon State University, Corvallis, OR 97331}

\author{C. Jess Riedel}

\address{Perimeter Institute for Theoretical Physics, Waterloo, Ontario N2L
2Y5, Canada}

\address{IBM Watson Research Center, Yorktown Heights, NY 10598}

\author{Wojciech H. Zurek}

\address{Theoretical Division, MS-B213, Los Alamos National Laboratory, Los
Alamos, NM 87545}
\begin{abstract}
Quantum Darwinism recognizes the role of the environment as a communication
channel: Decoherence can selectively amplify information about the
pointer states of a system of interest (preventing access to complementary
information about their superpositions) and can make records of this
information accessible to many observers. This redundancy explains
the emergence of objective, classical reality in our quantum Universe.
Here, we demonstrate that the amplification of information in realistic
spin environments can be quantified by the quantum Chernoff information,
which characterizes the distinguishability of partial records in individual
environment subsystems. We show that, except for a set of initial
states of measure zero, the environment always acquires redundant
information. Moreover, the Chernoff information captures the rich
behavior of amplification in both finite and infinite spin environments,
from quadratic growth of the redundancy to oscillatory behavior. These
results will considerably simplify experimental testing of quantum
Darwinism, e.g., using nitrogen vacancies in diamond.
\end{abstract}
\maketitle
Quantum Darwinism is a framework that allows one to understand the
emergence of the objective, classical world from within an ultimately
quantum Universe \cite{Zurek09-1,Zurek14-1}. Objects decohere in
the presence of their environment \cite{Joos03-1,Zurek03-1,Schlosshauer08-1},
resulting in a mixture of pointer states. During this process, the
environment acts as an amplification channel \cite{Zwolak14-1}, acquiring
and transmitting redundant information about the pointer states. This
occurs via the imprinting of the object's pointer observable \cite{Zurek81-1}
$\PB=\sum_{\hat{s}}\pi_{\hat{s}}\proj{\hat{s}}{\hat{s}}$ onto conditional
states, $\CSp=\bra{\hat{s}}\rho_{\S\F}\ket{\hat{s}}/p_{\hat{s}}$,
of fragments of the environment $\E$, where $\hat{s}=1,\ldots,D_{\S}$
label the pointer states, $p_{\hat{s}}$ are their probabilities,
and $\rho_{\S\F}$ is the joint state of the system $\S$ and some
fragment $\F$.

The quantum mutual information 
\[
\MI=H_{\S}+H_{\F}-H_{\S\F},
\]
where $H_{\mathcal{A}}=-\tr\rho_{\mathcal{A}}\log_{2}\rho_{\mathcal{A}}$
is the von Neumann entropy, quantifies the correlations generated
between the system and the fragment. As was recently shown, it is
a sum of classical and quantum components \cite{Zwolak13-1}, with
the former being the Holevo quantity \cite{Kholevo73-1,Nielsen00-1}
and the latter the quantum discord \cite{Zurek00-1,Ollivier02-1,Henderson01-1}.
The Holevo quantity \cite{Kholevo73-1}, 
\begin{equation}
\Holp=H\left(\sum_{\hat{s}}p_{\hat{s}}\CSp\right)-\sum_{\hat{s}}p_{\hat{s}}H\left(\CSp\right),\label{eq:Holevo}
\end{equation}
bounds the amount of classical information transmittable by a quantum
channel. In our case, the classical information is about the pointer
states of the system and the environment fragment's state is the output
of a quantum channel. Redundant records are available when many fragments
of $\E$ contain information about $\S$, i.e., when 
\begin{equation}
\avg{\Holp}_{\Fd}\cong\left(1-\delta\right)H_{\S},\label{eq:CondRed}
\end{equation}
with $\Fd\ll\Es$ the number of subsystems in the fragment $\F$ of
$\E$ needed to acquire $\left(1-\delta\right)H_{\S}$ bits of information.
Here, $H_{\S}=H\left(\PB\right)$ is the missing information about
$\S$, $\delta$ is the information deficit (the information the observers
are prepared to forgo), and $\Es$ is the total number of subsystems
in the environment. The average $\avg{\cdot}_{\Fd}$ is taken over
all choices of $\F$ with size $\Fd$.

The redundancy -- the number of records of the information -- is just
\begin{equation}
R_{\delta}=\Es/\Fd.\label{eq:Red}
\end{equation}
Redundancy allows many observers to independently access information
about a system \cite{Ollivier04-1,Blume06-1} and guarantees that
they will arrive at compatible conclusions about its state. The presence
of redundancy distinguishes the preferred quantum states (that can
aspire to classicality) from the overwhelming majority of possible
state in Hilbert space \cite{Dalvit01-1,Blume05-1,Riedel12-1} and
explains the emergence of objective classical reality in a quantum
universe. 
\begin{figure*}[!t]
\centering{}\includegraphics[width=7in]{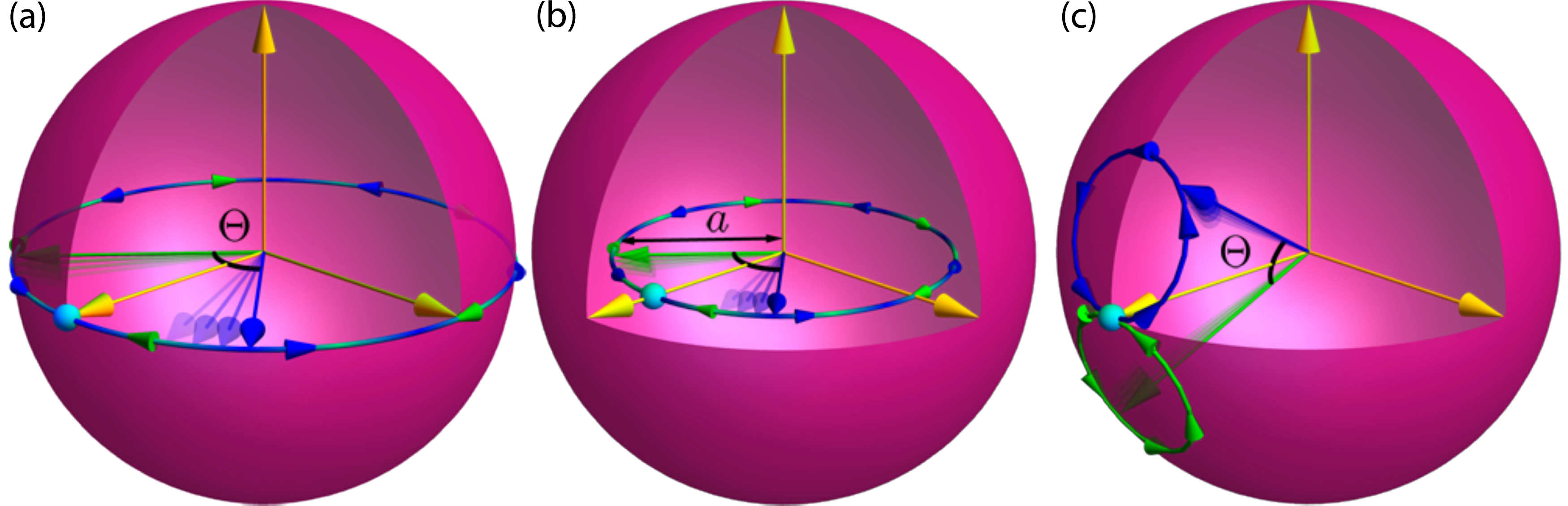}\protect\caption{Record acquisition by an environment spin: Three-dimensional trajectories
on the Bloch sphere that depict the acquisition of a record by a single
two-level (spin) environment subsystem. (a) A qubit system interacts
with a spin environment subsystem with $\omega_{k}=0$. (See Eq. \eqref{eq:Ham}.)
The two conditional states of an individual environment spin, $\rho_{k\left|\uparrow\right.}$
(green) and $\rho_{k\left|\downarrow\right.}$ (blue), rotate in opposite
directions from the initial state (light blue sphere) due to interaction
with the spin-up and spin-down pointer states of the system. For pure
states, the decoherence and completeness of the record is determined
by the angle $\Theta$ between the Bloch vectors for $\rho_{k\left|\uparrow\right.}$
and for $\rho_{k\left|\downarrow\right.}$ -- the angle that appears
in the quantum Chernoff bound (QCB), Eq. \eqref{eq:QCBspinESupp}.
(b) The same as (a), but with an initially mixed subsystem state.
The mixedness contracts the Bloch vectors (here, to a length $a=11/16$)
and reduces the ability of an environment spin to store distinguishable
records of the system's pointer states. (c) Same as (a) but with a
subsystem self-Hamiltonian, $\omega_{k}\sigma_{k}^{x}=\pi\sigma_{k}^{x}/2$.
The latter contribution to the Hamiltonian can enhance or reduce the
susceptibility of the subsystem to be rotated by $\protect\S$, depending
on the initial state, time, etc. While the case of $\omega_{k}=0$
gives analogous behavior to photons, the case of $\omega_{k}\protect\neq0$
is relevant for more general environments, such as the nuclear spin
environment of a nitrogen vacancy (NV) in diamond. \label{fig:Geo}}
\end{figure*}

\begin{figure*}
\begin{centering}
\includegraphics[width=7in]{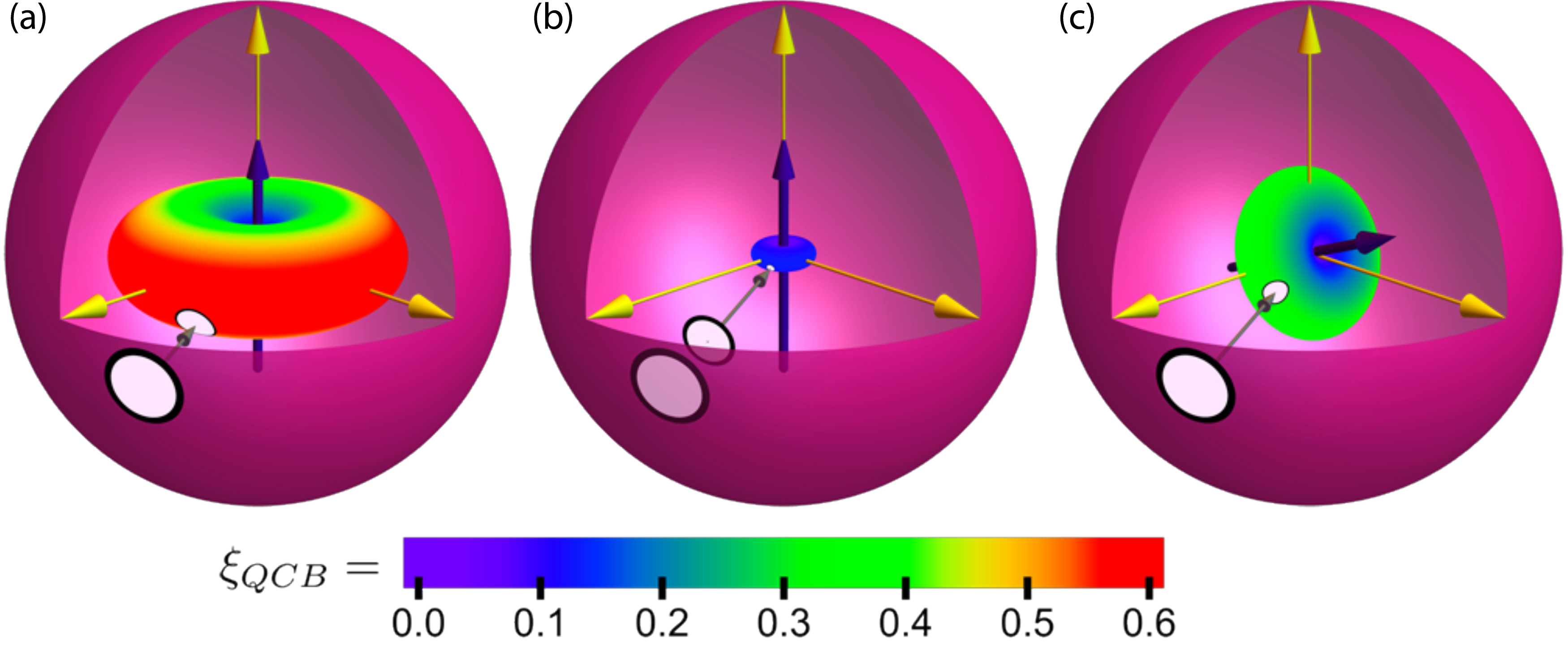} 
\par\end{centering}

\protect\caption{The contribution to the quantum Chernoff information, Eqs. \eqref{eq:TQCB}
and \eqref{eq:QCBspinESupp}, for a single environment spin $k$ with
(a) $\omega_{k}=0$ \& $a=1$, (b) $\omega_{k}=0$ \& $a=11/16$,
and (c) $\omega_{k}=\pi/2$ \& $a=1$. (In all cases, $t=15\pi/64$
and $g_{k}=1/2$.) These parameters are the same as in Fig. \ref{fig:Geo}.
The white patches map a region of initial states of the spin, specified
by ($a$, $\theta$, $\phi$) in the Bloch sphere, to a region ($\xi_{QCB}$,
$\theta$, $\phi$) of the central, toroidal structure. For the mixed
state case, two patches are shown: ($1$, $\theta$, $\phi$) in light
pink and ($a$, $\theta$, $\phi$) in white. These structures demonstrate
that there is only a single axis -- an ``insensitive axis'' shown
as a dark purple arrow -- of initial states that have no information
transferred to them, and, consequently, do not contribute to the redundancy.
When $\omega_{k}=0$, this axis is the $z$-axis -- these states of
the environment subsystem cannot decohere the system and have zero
susceptibility to acquire information. In a sense, they are the pointer
states of the environment subsystem with respect to decoherence induced
by the system. For $\omega_{k}\protect\neq0$, the insensitive axis
is time-dependent due to the intrinsic dynamics of the environment.
These structures show explicitly that redundancy is a universal feature
of pure decoherence models; the initial states that preclude the acquisition
of a partial record form a set of measure zero. In other words, essentially
all spins in the environment will be imprinted with a partial record
of the system's state. (See Fig. \ref{fig:Geo} for an illustration
of this process.) These partial records can be therefore investigated
experimentally by tomography of individual spins. \label{fig:LimitR}}
\end{figure*}

We note that, in the context of our everyday experience, $H_{\S}$
above is not the thermodynamic entropy of $\S$. Rather, it is usually
the missing information about the \emph{relevant} degrees of freedom
of $\S$. This is an important distinction: The thermodynamic entropy
of a cat, for instance, will vastly exceed the information the observer
is most interested in -- the one bit of greatest interest in the setting
imagined by \textcolor{black}{Schrödinger \cite{Schrodinger35-1}}.
Moreover, only such salient features of macrostates will usually be
preserved or will evolve in a predictable manner under the combined
influence of the self-Hamiltonian and of the decohering environment.
The condition for the preservation of macrostates under copying (or
under monitoring by, e.g., the environment -- the cause of decoherence)
is the orthogonality of the subspaces that support such macrostates
\cite{Zurek13-1}. It is reminiscent of the condition for preservation
of microstates under measurements \cite{Zurek07-3}, but the degeneracy
within macrostates allows for evolution and even for the change of
their thermodynamic entropy (which would certainly occur in the example
of the cat we have just invoked).

To illustrate quantum Darwinism, we examine a decohering qubit. In
this case, the Hilbert space of the qubit is of course too small to
allow for the above distinction, leaving no room for the thermodynamic
entropy. We take the interaction between the central qubit and the
environment to be the pure decoherence Hamiltonian 
\begin{equation}
\H=\sigma_{\S}^{z}\sum_{k=1}^{\Es}g_{k}\sigma_{k}^{z}+\sum_{k=1}^{\Es}\vec{\omega}_{k}\cdot\vec{\sigma}_{k},\label{eq:Ham}
\end{equation}
where the system's self-Hamiltonian is assumed to be negligible and
$k$ specifies an environment spin. The pointer observable, $\PB$,
is $\sigma_{\S}^{z}$. The self-Hamiltonian of the environment could
be due to a magnetic field, with $\omega_{k}$ the characteristic
frequency of the spin in that field. We will take $\vec{\omega}_{k}\cdot\vec{\sigma}_{k}=\omega_{k}\sigma_{k}^{x}$
for all specific expressions. The initial state is taken to be 
\begin{equation}
\rho\left(0\right)=\rho_{\S}\left(0\right)\otimes\left[\bigotimes_{k=1}^{\Es}\rho_{k}\left(0\right)\right].\label{eq:InitState}
\end{equation}
For part of this work, we will assume an environment with symmetric,
time-independent couplings and initial states ($g_{k}=g$, $\omega_{k}=\omega$,
$\rho_{k}\left(0\right)=\rho_{r}$ for all $k$). All expressions
will be generalized to non-symmetric states and non-symmetric (potentially,
time-dependent) couplings by taking a suitable average. In addition
to giving examples of environments that are non-i.i.d.\ (not independent
and identically distributed) and the possibility of visualizing the
acquisition of partial records, this class of spin environments is
the natural stepping-stone to finite, but higher dimensional, models
where the information transferred into the environment pertains to
coarse-grained observables of the system. Moreover, due to the prevalence
of experimentally characterizable spin systems, the models discussed
here will help test the underlying ideas of quantum Darwinism in a
laboratory setting using, for instance, nitrogen vacancy (NV) centers
\cite{Jelezko06-1,Doherty13-1,Schirhagl14-1}.

Starting from the initial product state, the system and the environment
will become correlated as they interact. Observers wanting to determine
the pointer state of the system, a $\sigma_{\S}^{z}$ eigenstate in
this case, need to distinguish the messages contained in the intercepted
fragment of $\E$. Evolution of the state, Eq. \eqref{eq:InitState},
generated by the Hamiltonian in Eq. \eqref{eq:Ham} will result in
the conditional states 
\[
\CSp=\bigotimes_{k\in\F}\rho_{k\left|\hat{s}\right.}
\]
for $\hat{s}=\up,\,\down$. The connection between the distinguishability
of these states and the Holevo quantity can be made via Fano's inequality
\cite{Nielsen00-1,Cover06-1}, 
\begin{equation}
\Holp\ge H_{\S}-H\left(P_{e}\right),\label{eq:Fano}
\end{equation}
where $P_{e}$ is the probability of error -- of incorrectly identifying
the conditional state. In the i.i.d.\ case, the quantum Chernoff
bound (QCB) shows that the error probability in discriminating the
two ``sources'' -- here, the pointer states -- decays as 
\[
P_{e}\sim\exp\left(-\Fs\xi_{QCB}\right)
\]
in the asymptotic regime \cite{Audenaert07-1,Audenaert08-1,Nussbaum09-1}.
The decay rate of the error is given by 
\[
\xi_{QCB}=-\ln\tr\left[\Cku^{c}\Ckd^{1-c}\right],
\]
which is the \emph{quantum Chernoff information}. The value of $c$
is that which maximizes this exponent and satisfies $0\le c\le1$.
The quantity $\xi_{QCB}$ gives a generalized measure of overlap between
two states.

In quantum Darwinism the QCB provides the measure of distinguishability,
including the case of non-i.i.d.\ environment components \cite{Zwolak14-1}.
Amplification of information about the pointer states is reflected
in the rapid decay of ignorance with the size of the fragment. Enforcing
the condition, Eq. \eqref{eq:CondRed}, together with Fano's inequality,
Eq. \eqref{eq:Fano}, gives an estimate for the redundancy \cite{Zwolak14-1}
\begin{equation}
R_{\delta}\simeq\Es\frac{\bar{\xi}_{QCB}}{\ln1/\delta}.\label{eq:QCBSupp}
\end{equation}
This estimate stems from a lower bound on the redundancy as $\delta\to0$
\cite{Zwolak14-1}. The quantity $\bar{\xi}_{QCB}$ is the ``typical''
quantum Chernoff information (for potentially non-i.i.d.\ environments),
\begin{equation}
\QCB=-\ln\avg{\tr\left[\Cku^{c}\Ckd^{1-c}\right]}_{k\in\E}.\label{eq:TQCB}
\end{equation}
It characterizes the distinguishability averaged over contributions
of \emph{individual} environment subsystems (i.e., fragments of size
1). Equation \eqref{eq:TQCB} can be maximized over $0\le c\le1$.
This may not always be practical. We shall see it can be done for
spin environments. (Previously, we demonstrated it can be done for
photon environments \cite{Zwolak14-1}.) Otherwise, though, the parameter
$c$ can just be set to some value between 0 and 1, e.g., $c=1/2$,
to get a weaker lower bound on the redundancy (and the quantum Chernoff
information). As seen from Eqs. \eqref{eq:QCBSupp} and \eqref{eq:TQCB},
the distinguishability (alternatively, the overlap) of the conditional
states, $\Cku$ and $\Ckd$, of the $k^{\mathrm{th}}$ environment
subsystem determines its contribution to the QCB and the redundancy.

Equation\ \eqref{eq:QCBSupp} is an extraordinarily practical tool
that we will exploit in this work: To calculate the redundancy (and,
hence, the amplification), one need only study \emph{individual} environment
subsystems, rather than states in the exponentially large Hilbert
space of the fragment. Moreover, \emph{the system's probabilities
appear only as small corrections to Eq.}\ \emph{\eqref{eq:QCBSupp}
}(except for the trivial case when $p_{\uparrow}$ is zero or one).
Thus, the quantity of interest is the typical quantum Chernoff information
$\bar{\xi}_{QCB}$, which quantifies the distinguishability of the
states of the environment. \emph{This shifts the focus from the objective
existence of the state of the system to the redundancy -- hence, accessibility
by many observers, the hallmark of objectivity -- of the records of
its state in the environment. The formation and redundancy of these
records depend on how the environment responds to the presence of
the system.}

The quantum Chernoff information allows one to arrive at useful estimates
of redundancy based on measurements of only single subsystems of the
environment rather than tomography of whole fragments $\F$, a task
that is exponentially difficult in $\Fs$. Although mathematical models
of decoherence sometimes make strong assumptions about the form of
the interaction Hamiltonian, the key prediction of these models --
large redundancies -- can be tested experimentally \emph{without}
such assumptions. Therefore, it is hoped that the results presented
here will enable testing of quantum Darwinism in the laboratory.

\section{Results}

Here we set the stage by using the quantum Chernoff information to
investigate the transfer of information between a qubit system and
spins of the environment. Specifically, we obtain general formulas
and then use them to analyze paradigmatic examples of spin environments.
We also elucidate the relation between the QCB, redundancy, and decoherence.

\subsection{Decoherence and Information}

Focusing on an individual spin from the environment, we now study
the relation between decoherence and the imprinting of a partial record.
According to Eq. \eqref{eq:Ham}, the interaction with the system
imprints its $\hat{s}^{\mathrm{th}}$ pointer state on the $k^{\mathrm{th}}$
subsystem, $\rho_{k\left|\hat{s}\right.}$, by a controlled unitary,
i.e., rotating it from its initial state to the state 
\[
\rho_{k\left|\hat{s}\right.}=\V_{\hat{s}}\rho_{k}\left(0\right)\V_{\hat{s}}^{\dagger},
\]
with 
\begin{align*}
\V_{\hat{s}} & =\bra{\hat{s}}\exp\left(-\im\H t\right)\ket{\hat{s}}.
\end{align*}
This process of controlled rotation is depicted in Fig. \ref{fig:Geo}.
The off-diagonal elements of $\rho_{\S}\left(t\right)$ will be suppressed
by the decoherence factor, $\gamma=\prod_{k\in\E}\gamma_{k}$, with
contributions 
\[
\gamma_{k}=\tr\left[\V_{\uparrow}\rho_{k}\left(0\right)\V_{\downarrow}^{\dagger}\right]
\]
from each subsystem $k$. Each contribution to the QCB is, ignoring
the logarithm, 
\[
\tr\left[\V_{\uparrow}\rho_{k}\left(0\right)^{c}\V_{\uparrow}^{\dg}\mathcal{V}_{\downarrow}\rho_{k}\left(0\right)^{1-c}\V_{\downarrow}^{\dagger}\right].
\]
For a pure initial state (or a purified state where the observer has
access to the purifying system), we thus have 
\begin{equation}
\bar{\xi}_{QCB}=-\ln\avg{\left|\gamma_{k}\right|^{2}}_{1}=-\ln\avg{\cos^{2}(\Theta/2)}_{1},\label{eq:QCB-Dec}
\end{equation}
where the second equality is for spin environments and the average
$\left\langle \cdot\right\rangle _{1}$ is over the individual components
of the environment, i.e., fragments of size $\Fs=1$. The angle $\Theta$
is how much the conditional states of the environment spin are separated
on the Bloch sphere, as shown in Fig. \ref{fig:Geo}. The redundancy
is therefore 
\begin{equation}
R_{\delta}\simeq\Es\frac{\ln\avg{\cos^{2}(\Theta/2)}_{1}}{\ln\delta}.\label{eq:Rpure}
\end{equation}
There is thus a direct correspondence between decoherence and redundancy
when the environment is initially pure -- when there is decoherence,
records of the system's pointer states will be proliferated into the
environment. The estimate of $R_{\delta}$ using Eq. \eqref{eq:QCBSupp}
comes from a lower bound on the redundancy. However, in the case of
an initially pure $\E$ state, Eq. \eqref{eq:QCBSupp}, and hence
Eq. \eqref{eq:Rpure}, is exact as $\delta\to0$. This is easily shown
by expansion of the mutual information, as shown in the Methods. Any
initial mixedness in the relevant subspace of the environment subsystem
-- in the space that acquires the record -- will decrease the information
about $\S$ observers can deduce, as we will now show.

\subsection{The Quantum Chernoff Information for Mixed Environments}

To find the QCB and $R_{\delta}$ for an initially mixed spin environment,
we examine $\tr\left[\V_{1}\rho_{k}\left(0\right)^{c}\V_{1}^{\dg}\mathcal{V}_{2}\rho_{k}\left(0\right)^{1-c}\V_{2}^{\dagger}\right]$,
which appears in Eq. \eqref{eq:TQCB}. Letting $\mathcal{D}=\mathcal{U}^{\dg}\V_{1}^{\dg}\mathcal{V}_{2}\mathcal{U}$
with $\mathcal{U}^{\dg}\rho_{k}\left(0\right)\mathcal{U}=\mathrm{Diag}\left[\lambda_{-},\lambda_{+}\right]$,
this quantity is given as 
\begin{eqnarray}
\tr\left[\left(\begin{array}{cc}
\lambda_{-}^{c} & 0\\
0 & \lambda_{+}^{c}
\end{array}\right)\mathcal{D}\left(\begin{array}{cc}
\lambda_{-}^{1-c} & 0\\
0 & \lambda_{+}^{1-c}
\end{array}\right)\mathcal{D}^{\dg}\right]\nonumber \\
=\left|\mathcal{D}_{11}\right|^{2}+\left|\mathcal{D}_{21}\right|^{2}\left(\lambda_{-}^{c}\lambda_{+}^{1-c}+\lambda_{-}^{1-c}\lambda_{+}^{c}\right).
\end{eqnarray}
This is symmetric about $c=1/2$ and attains its minimum there (and
thus it maximizes Eq. \eqref{eq:TQCB}). For higher dimensional environment
subsystems the minimum will not necessarily occur at $c=1/2$, even
for pure decoherence. Further, without loss of generality, let $\rho_{k}\left(0\right)$
be along the $z$-axis of the Bloch sphere and $\mathcal{D}$ be a
rotation about an axis in the $xy$-plane. One then obtains 
\begin{equation}
\bar{\xi}_{QCB}=-\ln\left[\avg{1-\MD\sin^{2}\Theta/2}_{1}\right].\label{eq:QCBspinESupp}
\end{equation}
The quantity 
\[
\MD=1-\sqrt{4\lambda_{-}\lambda_{+}}=1-\sqrt{1-a^{2}},
\]
where $a$ is the length of the Bloch vector, is a measure of the
mixedness of the state. Unless otherwise stated, the quantities $\MD$,
$\Theta$, etc., depend on the environment subsystem $k$.

{\em No External Field, $\omega_{k}=0$}: For $\omega_{k}=0$ in
the Hamiltonian \eqref{eq:Ham}, the QCB takes on the form 
\begin{equation}
\bar{\xi}_{QCB}=-\ln\left[\avg{1-\MD\sin^{2}2g_{k}t\,\sin^{2}\theta}_{1}\right],\label{eq:QCBw0}
\end{equation}
where $\theta$ is the angle of the initial environment spin state
from the $z$-axis. Notice that the polar angle, $\phi$, does not
appear in $\bar{\xi}_{QCB}$ when no external field is present, and
hence the QCB is rotationally symmetric about the $z$-axis. This
axis is special: It is insensitive to the state of the system, as
the $z$-states are eigenstates of the Hamiltonian and these states
have zero capacity to acquire information (in some sense, they are
the pointer states of the \emph{environment} \emph{spin} \cite{Zurek81-1}).

Figure \ref{fig:LimitR}(a,b) shows the QCB mapped from the Bloch
sphere for an initially pure and mixed environment spin. The QCB forms
a toroidal structure around the $z$-axis. Only along this axis is
the QCB zero. In other words, all the possible initial states of the
environment will proliferate redundant information, except ones that
have all subsystems initially in $z$-eigenstates or mixtures thereof.
Redundancy is thus inevitable, as these structures show.

{\em With an External Field, $\omega_{k}\neq0$}: For $\omega_{k}\sigma_{k}^{x}$
in the Hamiltonian \eqref{eq:Ham}, the quantum Chernoff information
comes out to be 
\begin{equation}
\bar{\xi}_{QCB}=-\ln\left[\left\langle 1-\MD f\left(t\right)\sin^{2}\tilde{\theta}\left(t\right)\right\rangle _{k}\right],\label{eq:GeneralQCB}
\end{equation}
where 
\begin{equation}
f\left(t\right)=\frac{\left[g_{k}^{4}\sin^{2}\left(2\tilde{\omega}_{k}t\right)+4g_{k}^{2}\omega_{k}^{2}\sin^{2}\left(\tilde{\omega}_{k}t\right)\right]}{\tilde{\omega}_{k}^{4}}.\label{eq:f}
\end{equation}
In this expression, we have used the effective field strength, $\tilde{\omega}_{k}=\sqrt{g_{k}^{2}+\omega_{k}^{2}}$,
felt by the environment spin. The factors in the average depend on
the initial state and parameters describing the $k^{\mathrm{th}}$
environment subsystem. For completeness, note that $\lambda=1-\sqrt{4\lambda_{-}\lambda_{+}}=1-\sqrt{1-a^{2}}$
is the same factor as in the case $\omega_{k}=0$. The angle $\tilde{\theta}\left(t\right)$
is the angle of the initial state of the environment subsystem from
the ``insensitive axis'' at time $t$, which is defined by the Bloch
angles $\theta^{\star}=\arctan\left[\pm\omega_{k}\tan\left(\tilde{\omega}_{k}t\right)/\tilde{\omega}_{k}\right]$
and $\phi^{\star}=\pi\mp\pi/2$ (i.e., it depends on the time and
parameters of the $k^{\mathrm{th}}$ environment subsystem's Hamiltonian).
When the environment subsystem initially points in the direction of
this axis, then at time $t$ it will contain no information and therefore
will have zero contribution to the redundancy. This (time-dependent)
axis is thus the counterpart to the $z$-axis when $\omega_{k}=0$.

Figures \ref{fig:Geo}(c) and \ref{fig:LimitR}(c) show the acquisition
of a record and $\bar{\xi}_{QCB}$ for $\omega=\pi/2$, respectively.
While the behavior is different \textendash{} the ``toroids'' rotate
\textendash{} one still gets redundancy. The shape is also still highly
symmetric. This is clear from the expression, Eq.\ \eqref{eq:GeneralQCB},
above: At any given time, the object is rotationally symmetric about
the axis defined by $\theta^{\star}$ and $\phi^{\star}$. Thus, Eq.\ \eqref{eq:GeneralQCB}
shows that the Hamiltonian defines a unique structure on the Bloch
sphere of individual environment spins. The spatial extent of the
structure determines the redundancy achievable by states on $\E$,
and there is always a zero point defined by the ``insensitive axis'',
around which the structure is also rotationally symmetric. Moreover,
this demonstrates that redundancy is not fragile in the sense of being
prohibited by self-Hamiltonians of the individual environment spins.
Although the field $\omega\sigma^{x}$ can interfere with the ability
of the environment to decohere the system, the structures in Fig.
\ref{fig:LimitR} show that fields can actually enhance the ability
of some states to both decohere the system and acquire information
about it. Of course, if the field is strong enough, the environment
spin's state rotates so fast that it decouples, so it will neither
decohere nor acquire information about the system (this can be seen
from Eq.\ \eqref{eq:f}, $\lim_{\omega\to\infty}f\left(t\right)=0$,
giving $\lim_{\omega\to\infty}R_{\delta}=0$). Again, this shows that
only in particular cases -- cases of measure zero -- can redundancy
vanish.

\subsection{Examples}

The general results presented above set the stage for a direct application
of the QCB to example spin environments. They display a variety of
behaviors for the redundancy and the acquisition of records about
the system's state. Here we will discuss natural spin environments
relevant to bringing quantum Darwinism into the lab.

\subsubsection{Gaussian Decoherence and $R_{\delta}\propto t^{2}$}

\label{sec:gaussian}

Spin/two-level environments typically arise in solid state systems
where a central system, such as another spin, interacts with many
environment spins/two-level systems with a bounded total energy. Decoherence
in this paradigmatic setting was studied in Refs. \cite{Zurek04-1,Cucchietti05-1,Zurek07-2},
where it was shown that this universally results in Gaussian decoherence.
In this situation, the Hamiltonian is Eq. \eqref{eq:Ham} and the
coupling constants are assumed to depend on the environment size as
\begin{equation}
g_{k}^{2}=\mathcal{G}^{2}/\Es,\label{eq:ginvE}
\end{equation}
with $\mathcal{G}$ implicitly dependent on $k$. We note that the
coupling constants do not actually need to depend on environment size
to get Gaussian behavior. However, this dependence is physical (and
also makes a short time approximation unnecessary). We will set $\omega_{k}=0$
for convenience. The decoherence factor of the system decays as 
\begin{equation}
\left|\gamma\right|^{2}\sim\exp\left[-t^{2}/\tau_{D}^{2}\right],\label{eq:GaussDec}
\end{equation}
where $\tau_{D}$ is the decoherence time. This can be derived using
already demonstrated results.

Taking the relation between the QCB and the decoherence factor \emph{for
pure states}, Eq. \eqref{eq:QCB-Dec}, we can write 
\begin{equation}
\left|\gamma\right|^{2}\simeq\left\langle \left|\gamma_{k}\right|^{2}\right\rangle _{1}^{\Es}=e^{-\QCB\Es},\label{eq:DecFac}
\end{equation}
where the first approximate equality becomes exact in the limit $\Es\to\infty$
since $\left|\gamma_{k}\right|^{2}$ approaches $1$ with corrections
that can be written in a series in $1/\Es$. The $\QCB$ is given
by Eq. \eqref{eq:QCBw0} by letting the environment be pure ($\lambda=1$),
\begin{align*}
\bar{\xi}_{QCB} & =-\ln\left[\avg{1-\sin^{2}\left(2g_{k}t\right)\,\sin^{2}\theta}_{1}\right]\\
 & \approx\avg{4\mathcal{G}^{2}t^{2}\sin^{2}\theta}_{1}/\Es,
\end{align*}
where we have used Eq. \eqref{eq:ginvE} and assumed that $\Es$ is
very large. Again, all environment parameters implicitly depend on
$k$. Equation \eqref{eq:DecFac} now becomes Eq. \eqref{eq:GaussDec}
with 
\[
\frac{1}{\tau_{D}^{2}}=\avg{4\mathcal{G}^{2}\sin^{2}\theta}_{1}.
\]
This is the same result as in Refs. \cite{Zurek04-1,Cucchietti05-1,Zurek07-2}
taking into account the different definitions (The coupling $g_{k}$
in the Hamiltonian is defined in Refs. \cite{Zurek04-1,Cucchietti05-1,Zurek07-2}
as $g_{k}/2$ and $\sin^{2}\theta=4\sin^{2}\left(\theta/2\right)\cos^{2}\left(\theta/2\right)=4\left|\alpha\right|^{2}\left|\beta\right|^{2}$,
with implicit $k$ dependence). Note that for a mixed state environment
with the same $\avg{4\mathcal{G}^{2}\sin^{2}\theta}_{1}$ , the decoherence
time will be the same, i.e., the decoherence -- in contrast to amplification
and, hence, redundancy -- is independent of the mixedness of the environment.
\begin{figure}
\centering{}\includegraphics[width=8cm]{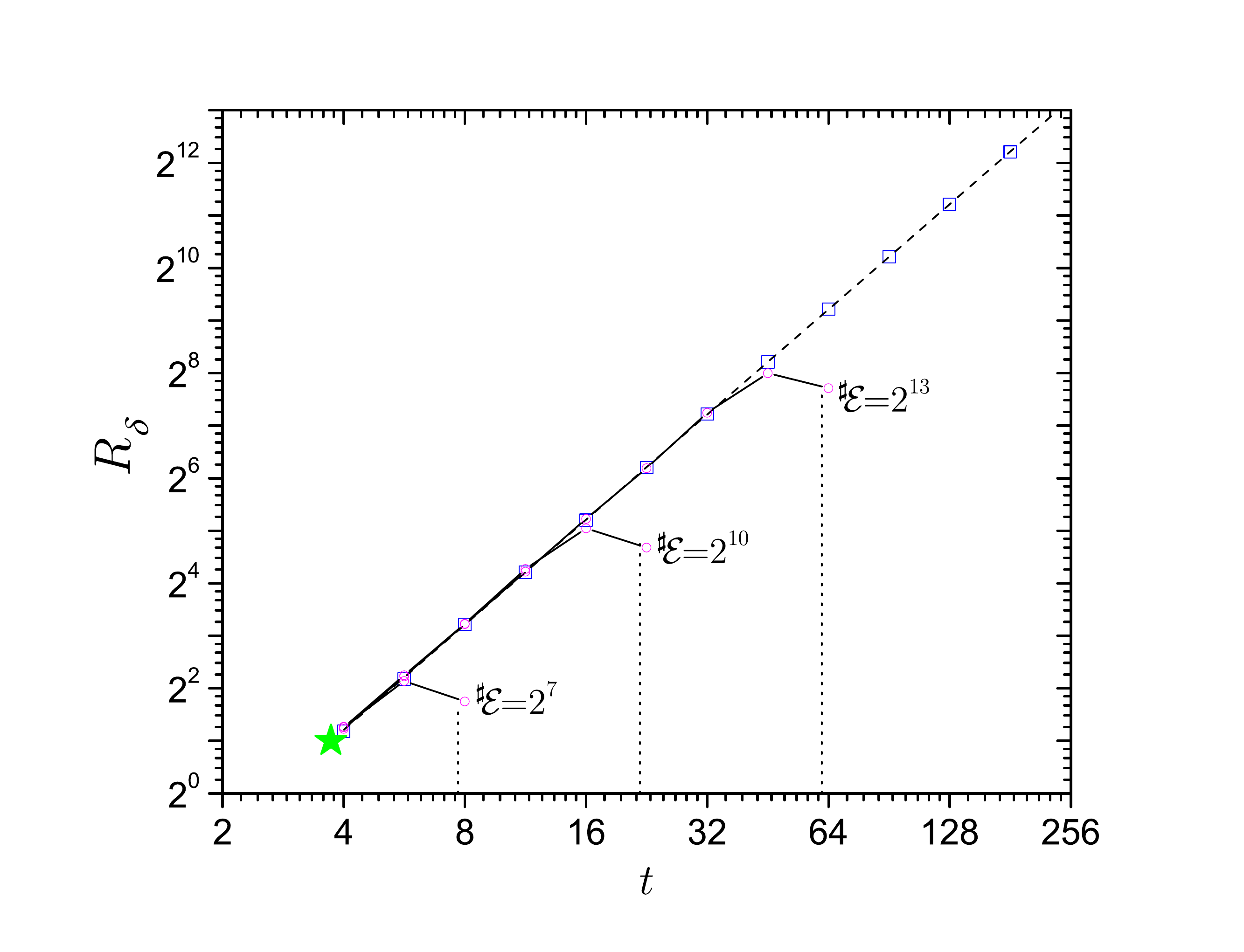}\protect\caption{The redundancy vs.\ time for Gaussian decoherence with an initially
pure environment. Here, the couplings $\mathcal{G}$ are chosen uniformly
from the interval $\left[-2,\,2\right]$, which gives $\tau_{D}=\sqrt{3}/4$.
The other parameters are $p_{\uparrow}=1/2$ and $\delta=10^{-16}$.
The blue squares are from computing the Holevo quantity with very
large environments and the black line is the QCB result, Eq. \eqref{eq:RQuad}.
The green star is the redundancy onset time, Eq. \eqref{eq:OnsetR}.
The three black dashed lines are for different finite $\protect\Es$
(for a fixed realization of the random coupling constants). For finite
environments, there is a quadratic growth of the redundancy up the
time, $\pi/4\sqrt{\protect\avg{g_{k}^{2}}}_{1}\sim\sqrt{\protect\Es}$,
where there is a partial flow of information from the environment
back into the system. This time for each of the finite environments
is indicated by the black dotted lines. Thus, even for finite environments,
one can expect to see signatures of Gaussian behavior in the redundancy.
\label{fig:OnsetAndGrowth}}
\end{figure}

\begin{figure}
\begin{centering}
\includegraphics[width=8cm]{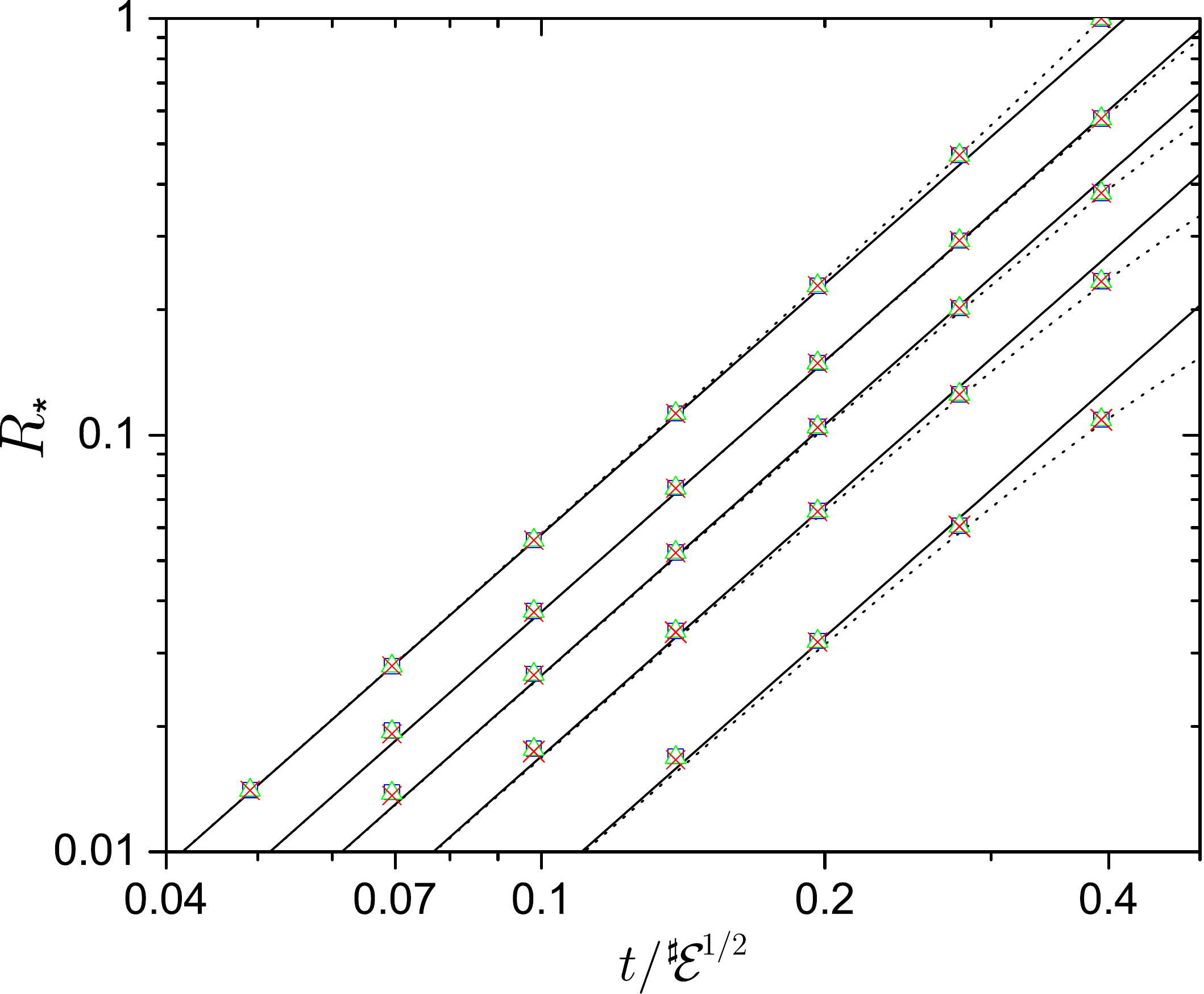} 
\par\end{centering}

\protect\caption{The amplification efficiency quantified by the quantum Chernoff information.
Here, the amplification is taken relative to a reference case, $R_{\star}\equiv\lim_{\delta\to0}R_{\delta}/R_{\delta}^{\protect\p}$.
The QCB predicts that the relative efficiency is $\xi_{QCB}/\xi_{QCB}^{\protect\p}$
with $\xi_{QCB}$ from Eq. \eqref{eq:QCBspinESupp} (black, dotted
lines). The quadratic growth of the redundancy, given by Eq. \eqref{eq:RQuad},
is shown as black, solid lines. For all data, the QCB prediction and
quadratic growth match well with the numerically computed results.
For very small relative efficiencies there is some deviation, which
is due to the finite $\protect\Fd$ obtainable numerically. The full
QCB result deviates from quadratic behavior for very long times (i.e.,
on the order of the recurrence time of an individual spin interaction,
which goes as $\pi/(4g_{k})\sim\sqrt{\protect\Es}$). The data in
the figure is as follows: The five lines are for varying haziness
(the initial entropy $h=H\left[\left(1+a\right)/2\right]$ of a single
environment spin \cite{Zwolak09-1,Zwolak10-1}) $h=0,\,1/5,\,2/5,\,3/5,$
and $4/5$ from top to bottom. Each set of symbols shows the numerical
result for $R_{\star}$ for three initially pure states of the system
($p_{\protect\up}=\protect\bra{\uparrow}\rho_{\protect\S}\left(0\right)\protect\ket{\uparrow}=1/2$,
$1/8$, and $1/32$) relative to the QCB reference case with with
$h=0$ and $t=\pi/8$. Further details can be found in the Methods.\label{fig:RQuad}}
\end{figure}

For the redundancy, including arbitrary mixedness, Equation \eqref{eq:QCBSupp}
becomes 
\[
R_{\delta}\simeq\Es\frac{\bar{\xi}_{QCB}}{\ln1/\delta}\approx\frac{\avg{4\lambda\mathcal{G}^{2}\sin^{2}\theta}_{1}t^{2}}{\ln1/\delta}.
\]
Assuming that the distributions of $\lambda$ and the $\mathcal{G}^{2}\sin^{2}\theta$
are independent, 
\begin{equation}
R_{\delta}\approx\frac{\avg{\lambda}_{1}t^{2}/\tau_{D}^{2}}{\ln1/\delta}\equiv\frac{\alpha t^{2}/\tau_{D}^{2}}{\ln1/\delta},\label{eq:RQuad}
\end{equation}
where $\alpha=\avg{\lambda}_{1}$ is a spin analog of receptivity
(generally, receptivity is just a proportionality factor between the
redundancy and the decoherence factor, $\alpha=R_{\delta}\log\left(\delta\right)/\log\left(\left|\gamma\right|^{2}\right)$,
see below). This indicates that redundancy grows quadratically with
time. Moreover, for $\Es\to\infty$, redundancy grows indefinitely
and without bound, even though the interaction energy between the
system spin and the infinite environment is bounded. Finite environments
have similar behavior except there is a time (determined by the inverse
of the typical interaction strength between the system and an environment
spin), beyond which information partially flows back into the system.
Figure \ref{fig:OnsetAndGrowth} shows the growth of redundancy with
time for both finite and (effectively) infinite environments. Figure
\ref{fig:RQuad} shows the redundancy for initially mixed environments
plotted with the results found using numerically exact techniques
for finite $\Fd$. This figure shows that the exact results approach
the QCB as $\delta$ gets smaller ($\Fd$ gets bigger).

In addition to the long time behavior for very large $\Es$, we can
also determine when redundant records start to form. From Eq. \eqref{eq:RQuad},
the onset of redundancy, $R_{\delta}\sim2$, happens at 
\begin{equation}
t^{\star}=\tau_{D}\sqrt{2\ln1/\delta}.\label{eq:OnsetR}
\end{equation}
That is, it is essentially the decoherence time multiplied by a factor
weakly dependent on the information deficit $\delta$. The latter
is order one for a large range of information deficits $\delta$.
Figure \ref{fig:OnsetAndGrowth} marks this onset time with a green
star.

Before we discuss another example, we note that the quadratic growth
in the redundancy is in sharp contrast to its linear growth for photon
environments \cite{Riedel10-1,Riedel11-1,Zwolak14-1,Korbicz14-1}.
Photons (or photon-like environments) are also amenable to calculations
using the QCB \cite{Zwolak14-1}. In this case, the redundancy grows
as 
\begin{equation}
R_{\delta}\simeq\frac{\alpha t/\tau_{D}}{\ln1/\delta},\label{eq:PhotonR}
\end{equation}
where $\alpha$ is the receptivity of the environment to making records,
which is a dimensionless quantity determined by the mixedness and
angular distribution of the incoming photon states \cite{Riedel10-1,Riedel11-1,Zwolak14-1,Korbicz14-1}.
The growth of the redundancy for photons is due to the linear increase
of the environment size in time, with each individual environment
component (each photon) acquiring a partial record of fixed fidelity,
at least on average. 

This is in contrast to the spin environment which has a fixed size
and continuously interacts with the system. The quadratic growth in
redundancy for spins is due to the increasing fidelity of the partial
records with time. We will show elsewhere \cite{Zwolak15-1} that
a flux of spins can represent the same acquisition of information
as photon environments -- and thus be used as a stand-in for photons. 

We note that there is a receptivity for both spins and photons. However,
for spins it is simply $\alpha=\avg{\lambda}_{1}$. Other factors
that affect the ability of the environment spins to acquire information
also affect their ability to decohere the system. This different form
of the receptivity is due to the fact that we have not made the equivalent
of a ``weak scattering'' approximation, but rather have only made
a weak coupling approximation for each environment spin and have allowed
each spin to continuously interact with the system.

\subsubsection{Other Non-i.i.d.\ Environments}

\begin{figure}
\begin{centering}
\includegraphics[width=8cm]{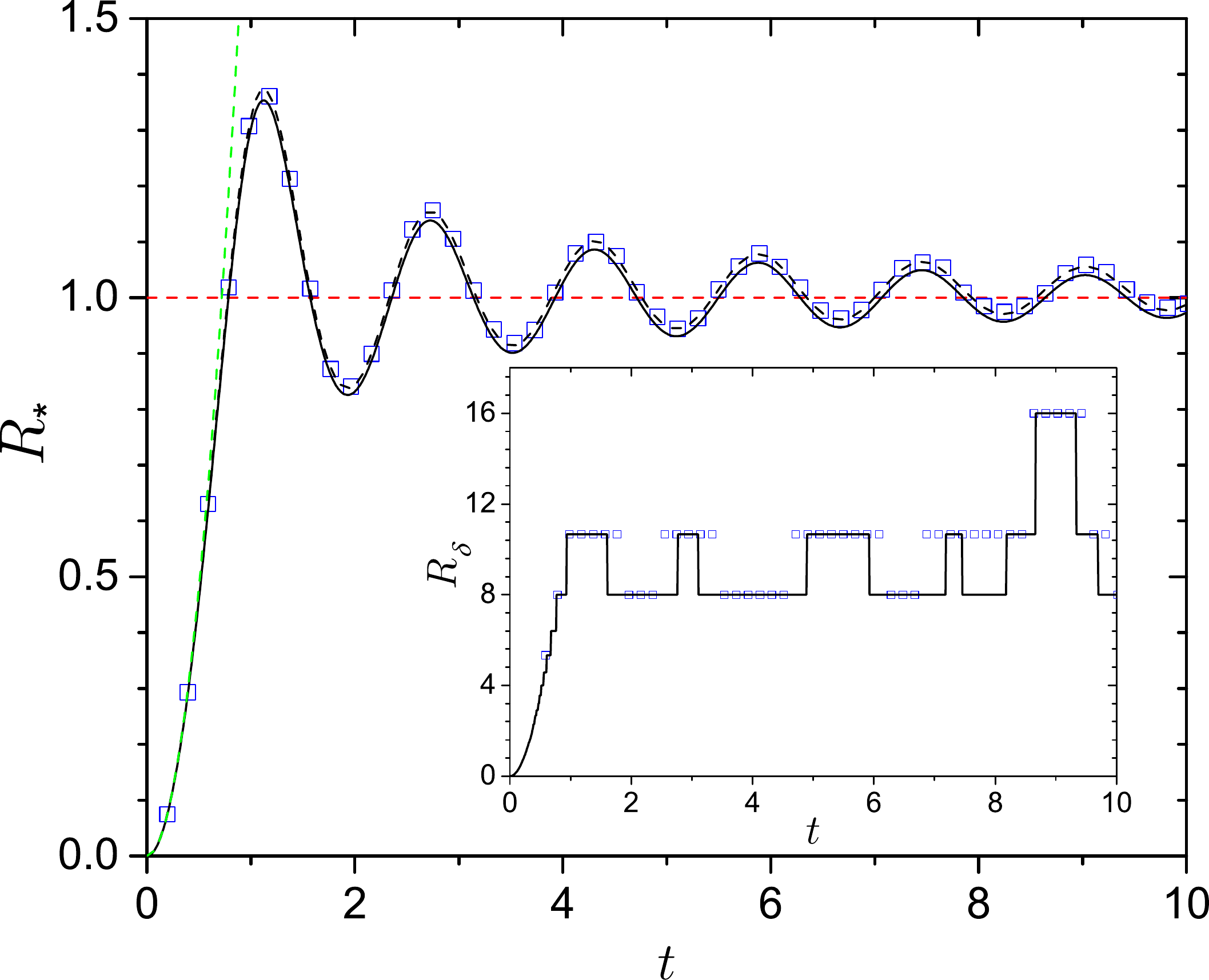} 
\par\end{centering}

\protect\caption{The amplification efficiency quantified by the quantum Chernoff information.
As with Fig. \ref{fig:RQuad}, the amplification is taken relative
to a reference case, $R_{\star}\equiv\lim_{\delta\to0}R_{\delta}/R_{\delta}^{\protect\p}$.
The plot shows the numerical data (blue squares), QCB (black solid
line), Gaussian regime (green dashed line), and an approximation that
has corrections for finite $\delta$ (black dashed line). The red
dashed line is the $t\to\infty$ result. When the coupling constants
$g_{k}$ come from a band of energies, $\left[0,1\right]$, the efficiency
of amplification initially increases quadratically with time -- i.e.,
it is in the universally present Gaussian regime -- and then develops
into an oscillatory behavior. The oscillations appear due to drawing
the coupling constants from a finite band. The inset shows $R_{\delta}$
for $\protect\Es=32$ and $\delta=10^{-1}$ for a \emph{single} set
of spins with random coupling constants drawn from $\left[0,1\right]$.
The exact numerical data, the open blue squares, shows that the oscillatory
features are still present even for this small environment. Moreover,
the black line shows a discretized application of the QCB. This shows
that in potential experiments with a very limited number of subsystems
of the environment can still display intricate dynamics of the redundancy
and the emergence of objective information. Moreover, the QCB can
capture this behavior and thus eliminate the need for a full tomographic
characterization of the system and environment. The Methods section
gives details of the data in the figure. \label{fig:Osc} }
\end{figure}

The Gaussian decoherence case above is not the only possible realization
of a central spin continually interacting with a fixed set of spins.
When a large number of environment spins couple strongly to the system,
then one can have still different behavior. Consider, for instance,
the Hamiltonian in Eq. \eqref{eq:Ham} with $\omega_{k}=0$ and the
coupling constants $g_{k}$ randomly drawn from the interval $\left[0,W\right]$,
with $W$ the ``bandwidth''. The QCB result, Eq. \eqref{eq:QCBSupp},
readily yields 
\begin{equation}
R_{\delta}\simeq\Es\frac{\ln\left[1-\frac{\lambda\sin^{2}\theta}{2}+\frac{\lambda\sin^{2}4Wt\,\sin^{2}\theta}{8Wt}\right]}{\ln\delta}\label{eq:QCBRandCoup}
\end{equation}
by averaging $1-\MD\sin^{2}2g_{k}t\,\sin^{2}\theta$ over the coupling
constants assuming that $\lambda$ and $\theta$ are constant. Figure
\ref{fig:Osc} shows the QCB result for the redundancy compared with
the results from averaging the Holevo quantity. As $\delta$ is decreased,
the numerical results converge to the QCB result. Notice that the
environment spins have a band of energies, which is responsible for
both the oscillation frequency and decay.

In this setting -- environment spins strongly coupled to the system
-- $\E$ will often have only a finite number of spins, i.e., $\Es$
can not be taken to be infinite. The qualitative features given by
the QCB, however, will still be present in the redundancy even for
relatively small number of spins in the environment, as shown in the
inset of Fig. \ref{fig:Osc}. Therefore, one can expect that in some
settings, more intricate dynamics of the redundancy will be present.

\section{Discussion}

The QCB demonstrates that redundancy is inevitable under pure decoherence:
The only way to avoid it for a spin environment is for all spins to
be in a completely mixed state (i.e., $a=0$, implying $\MD=0$) or
for all spins to be precisely aligned with the insensitive axis (i.e.,
$\Theta=0$). Figure \ref{fig:LimitR} shows this graphically. Pure
decoherence always gives rise to the redundant proliferation of information
except in rare -- measure zero -- cases. Furthermore, we showed that
the redundancy using the QCB estimate, Eq. \eqref{eq:QCBSupp}, agrees
with numerically exact results, which covers a wide variety of behavior
from Gaussian decoherence to oscillations. We discuss further examples
in a forthcoming publication \cite{Zwolak15-1}.

Although unavoidable imperfections ensure that real-world systems
rarely undergo pure decoherence, Eq. \eqref{eq:Ham}, models like
the one in Ref.~\cite{Riedel12-1} show that redundancy emerges even
in the presence of other types of environmental interactions. The
results presented here will help shed light on experiments where decoherence
and amplification are expected to occur for spins, such as NV-centers
immersed in an environment of nuclear spins. While the QCB gives the
exact asymptotic redundancy for the models here, two key features
of the estimate, Eq. \eqref{eq:QCBSupp}, are that (a) it does not
rely on idealized initial states or Hamiltonians when considered as
a lower bound and (b) it can be computed using only individual spin
measurements, with no need for complicated multipartite tomography.
Thus, whether a system self-Hamiltonian is present or not, and whether
there are more complicated interactions, one can demonstrate information
transfer into the environment with experimentally feasible measurements.
Our results, especially when confirmed experimentally, further elucidate
the acquisition of information by the environment and show why perception
of a classical objective reality in our quantum Universe is inescapable
\cite{Zwolak14-1,Brandao15-1}.

\section{Methods}

The numerical computation in Fig. \ref{fig:OnsetAndGrowth} are for
a symmetric environment with $\theta=\pi/2$ and $g_{k}=1$ for all
$k$ in order to make use of the method of Refs. \cite{Zwolak09-1,Zwolak10-1}.
Since $g_{k}=1$ for all $k$, the decoherence time is $\tau_{D}=1/2$,
which when rescaled by $1/\sqrt{\Es}$ will be zero for all practical
purposes (and well off the scale of the figure). We note that $R_{\star}$
is found numerically by fitting the decay of $H_{\S}-\Holp$ in order
to more rapidly approach the $\Fd\to\infty$ ($\delta\to0$) limit.
The numerical technique is exact for the computation of the Holevo
quantity, $\Hol$, for finite $\Fd$. Even though the QCB is the asymptotic
result ($\Fd\to\infty$), the numerical and analytical data match. 

The data for Fig. \ref{fig:Osc} are calculated as follows: The reference
case is evaluated at $t\to\infty$. The environment is taken to be
pure. The numerical data (open blue squares) was found by sampling
the random distribution of spins $10^{8}$ times to obtain $\delta$
as a function of $\Fs$. The solid black curve is the QCB, Eq. \eqref{eq:QCBRandCoup}
and the dashed black curve is from the expansion of the Holevo quantity
(see the following paragraph below). The dashed green curve is Gaussian
decoherence regime, giving $R_{\star}\propto t^{2}$. The information
deficit, $\delta=10^{-10}$, is still finite and thus there is a gap
between the QCB and the exact results, which closes as $\delta$ becomes
smaller. Note that the actual redundancy is quite large, linear in
the environment size. For the inset, the numerical data was evaluated
using an exact average of the Holevo quantity over all subsets of
size $\Fd$ of the $\Es=32$ spins for a fixed realization of the
random coupling constants. The discretized application of the QCB
takes $R_{\delta}=\Es/\lceil\Fs_{\delta}^{c}\rceil$, where $\Fs_{\delta}^{c}=\ln\delta/\ln\left[\avg{1-\MD\sin^{2}2g_{k}t\,\sin^{2}\theta}_{1}\right]$
is a continuous $\Fd$ and where the average is over the single realization
of random coupling constants. The redundancy jumps between discrete
steps since $\Fd$ takes on integer values, e.g., $R_{\delta}=16$
is for $\Fd=2$ and $R_{\delta}=32/3\approx10$ for $\Fd=3$, etc. 

For a pure system and environment, and a decoherence process due to
the Hamiltonian in Eq. \eqref{eq:Ham}, the mutual information is
given by $\MI=\EgF+\left[\EgE-\EgEF\right]$ \cite{Zwolak10-1}, with
the term in square brackets being the quantum discord \cite{Zwolak10-1,Zwolak13-1},
and the Holevo quantity by $\Hol=\EgF$. Here, the entropies are the
entropy of the system only decohered by some component of the environment,
either $\F$, $\E$, or $\E/\F$. These expressions can both be generalized
to the case of the system being mixed, see Eq. (A13) of Ref. \cite{Zwolak10-1}.
Expanding $\EgF$, inputing it into Eq. \eqref{eq:CondRed}, and using
that $\Es\gg\Fd$ -- so that each of the spins in the fragment can
be treated independently in the average -- gives $R_{\delta}=\Es\ln\avg{\cos^{2}\Theta/2}_{1}/\left(\ln\delta+\ln C\right)$
with $C=H_{\S}\left(p_{\uparrow}-p_{\downarrow}\right)\ln2/\left(p_{\uparrow}p_{\downarrow}\ln p_{\uparrow}/p_{\downarrow}\right)$.
$C\to\ln4$ for $p_{\uparrow}=p_{\downarrow}=1/2$. This equation
is, for any practical $\delta$, exact and shows that the redundancy
approaches the QCB result from above as $\delta\to0$. This equation
is the black dashed line in Fig. \ref{fig:Osc}.

\subsection*{Acknowledgments}

We would like to thank Salomon for color-scheme inspiration and the
Center for Integrated Quantum Science and Technology (IQST) and the
University of Ulm, where part of this work was carried out. Research
at the Perimeter Institute is supported by the Government of Canada
through Industry Canada and by the Province of Ontario through the
Ministry of Research and Innovation. This research was supported in
part by the U.S. Department of Energy through the LANL/LDRD Program
and, in part, by the John Templeton Foundation and the Foundational
Questions Institute Grant No. 2015-144057 on \textquotedblleft Physics
of What Happens.''

\subsection*{Author Contributions}

M.Z. developed the quantum Chernoff approach and performed calculations.
W.H.Z. helped define the project. All authors clarified and analyzed
the results and prepared the manuscript.

\end{document}